\documentclass[sigconf]{acmart}
\AtBeginDocument{%
  }

\copyrightyear{2025}
\acmYear{2025}
\setcopyright{cc}
\setcctype{by}
\acmConference[MM '25]{Proceedings of the 33rd ACM International Conference
on Multimedia}{October 27--31, 2025}{Dublin, Ireland}
\acmBooktitle{Proceedings of the 33rd ACM International Conference on
Multimedia (MM '25), October 27--31, 2025, Dublin,
Ireland}\acmDOI{10.1145/3746027.3759199}
\acmISBN{979-8-4007-2035-2/2025/10}

\newcommand{\relpose}[2]{\hat{\rho}_{#1,#2}}

\newcommand{\ha}[1]{\textcolor{blue}{#1}}

\usepackage{balance}

\newcommand{\m}[1]{\textcolor{blue}{#1}}

\begin{document}

\title{Reproducibility Companion Paper: Swarical: An Integrated Hierarchical Approach to Localizing Flying Light Specks}

 \author{Hamed Alimohammadzadeh}
 \email{halimoha@usc.edu}
 \orcid{0000-0003-2613-5010}
 \affiliation{%
   \institution{University of Southern California}
   \city{Los Angeles}
   \state{CA}
   \country{USA}
 }

 \author{Shahram Ghandeharizadeh}
 \email{shahram@usc.edu}
 \orcid{0000-0002-1792-7879}
 \affiliation{%
   \institution{University of Southern California}
   \city{Los Angeles}
   \state{CA}
   \country{USA}
 }
 
 \author{Federico Cunico}
 \email{federico.cunico@univr.it}
 \orcid{0000-0001-9619-9656}
 \affiliation{%
   \institution{University of Verona}
   \city{Verona}
   % \state{}
   \country{Italy}
 }

 \author{Joshua Springer}
 \email{joshua19@ru.is}
 \orcid{0000-0003-0137-1770}
 \affiliation{%
   \institution{Reykjavik University}
   \city{Reykjavik}
   % \state{}
   \country{Iceland}
 }


\begin{abstract}
This companion paper provides artifacts and instructions on replicating the experiments in the ACM Multimedia 2024 paper entitled "Swarical: An Integrated Hierarchical Approach to Localizing Flying Light Specks." \underline{Swar}m-based hierarch\underline{ical}, Swarical, is a localization technique that enables miniature drones, Flying Light Specks (FLSs), to accurately and efficiently localize and illuminate complex 2D and 3D shapes. It consists of two components, an offline planner and an online localization technique that executes on an FLS. The offline planner uses the FLS sensor specification for positioning to convert mesh files into swarms of FLSs. Some FLSs are dark and used only for localization. We reported the online localization technique to be fast and highly accurate. We describe how to reproduce this finding using our artifacts.
\end{abstract}



% %% A "teaser" image appears between the author and affiliation
% %% information and the body of the document, and typically spans the


\maketitle

\section{Introduction}\label{sec:intro}
Swarical is designed for a swarm of Flying Light Specks (FLSs) to illuminate shapes~\cite{shahram2021,shahram2022,circular2024,reliability2024,integrate2023,integrate2025,standbyfls2025,mmsys2023,imeta2023,flshaptics2023,haptics2024, flightpatterns2023, flshaptic23, shahram2022b, cmpfls2023, decentralized2023,uavmm2025,mcgeKeynote2025}.
An FLS is a small drone with some storage, processing power, networking capability to communicate with other FLSs, and sensors to compute its position relative to another FLS~\cite{dv2023,swarical2024,swarmer2023,swazure2024,shahram2022}. 
Our artifacts include complete software implementations, scripts, datasets, and instructions to reproduce the experimental results presented in~\cite{swarical2024}.
Swarical is designed to work with any device that enables an FLS to compute its position relative to another FLS.
We present one such device in~\cite{swarical2024}, namely, a small inexpensive Raspberry Pi Camera Module 3.
Hence, two separate GitHub repositories contain our artifacts:
\begin{itemize}
    \item Swarical,  \ha{\url{https://github.com/flyinglightspeck/Swarical}}.

    \item Position estimation using ArUco markers and Raspberry Pi Camera Module 3, \ha{\url{https://github.com/flyinglightspeck/aruco-pose-estimation}}.
\end{itemize}
The next two sections describe each repository in turn, presenting how to reproduce the results of~\cite{swarical2024}.

%\subsection{Summary of Swarical}
%In the original paper\cite{swarical2024}, we proposed Swarical, a Swarm-based hierarchical localization technique. It enables tiny drones, called Flying Light Specks (FLSs), to localize and illuminate 2D and 3D shapes. Swarical converts mesh files to point clouds using its offline component, called planner and emulates flight, position tracking, and localization of FLSs using its online part, called decentralized localization. The position tracking relies on an implementation using Raspberry cameras and ArUco markers. We proposed three variants of the decentralized localization techniques and compared the superior against SwarMer\cite{swarmer2023}. Results show that Swarical is more than 2x faster than SwarMer.

%\subsection{Artifacts Description}
% In addition, we provide small-scale experiments that are easier to replicate (described in Section \ref{sec:small_exp}). These artifacts are available at \\ \texttt{\small https://github.com/flyinglightspeck/Swarical} publicly. Additionally, the software used to conduct position tracking experiments is available in a separate repository at \\ \texttt{\small https://github.com/flyinglightspeck/aruco-pose-estimation}.

\section{Swarical}

\begin{figure}[ht]
\centering
\includegraphics[width=\columnwidth]
{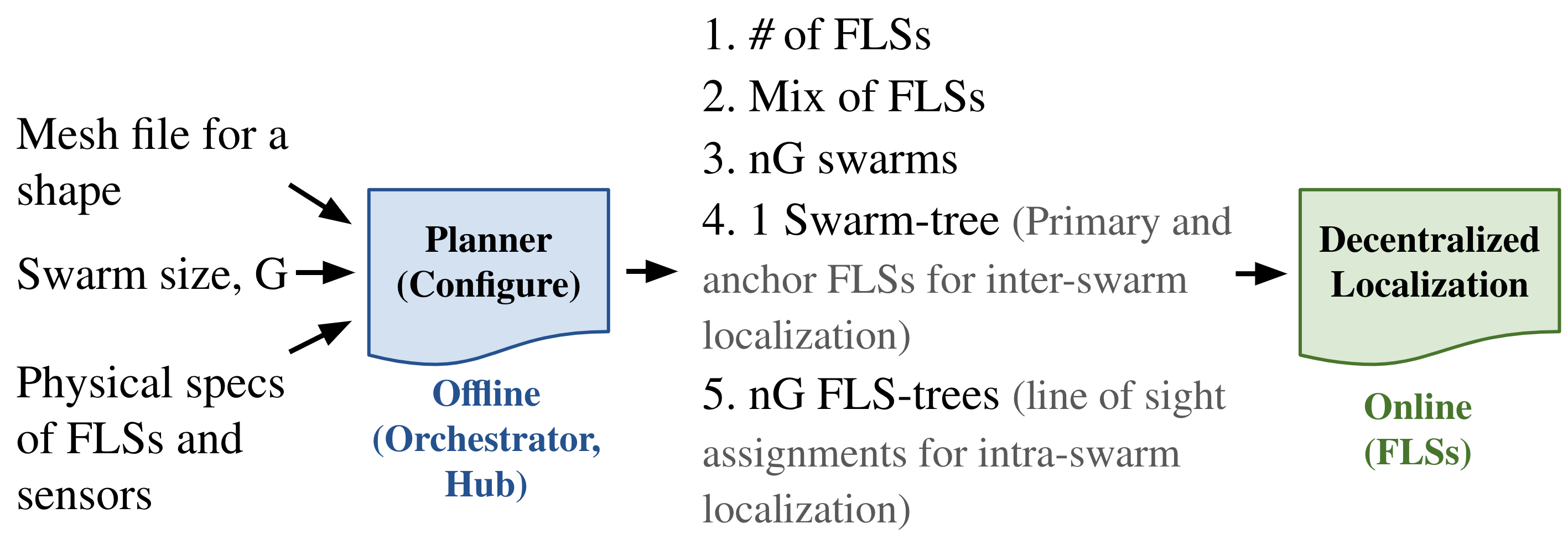}
\caption{Swarical, a divide-and-conquer framework.}
\label{fig:planner}
\end{figure}

Figure~\ref{fig:planner} shows the offline and online components of Swarical. The planner is the offline component. Its input is a mesh file for a shape, swarm size $G$, and the physical characteristics of each FLS, including its sensor, to estimate its position relative to another FLS.
Its output is the number of FLSs $F$, a mix of FLSs, the number of groups $nG$, a swarm-tree, and $nG$ FLS-trees.

The execution of the Swarical localization technique by an FLS is the online component.
The input to each FLS is:
\texttt{fls\_id}, \texttt{swarm\_id}, \texttt{anchor\_id}, \texttt{children\_ids}, and \texttt{ground\_truth\_location}.
We emulate an FLS as a software process.
A deployment consists of $F$ processes as computed by the planner.
Typically, we run these processes on a cluster of servers with one process (FLS) per CPU core.
Hence, a large experiment with $F$=1000 FLSs requires 1000 cores.
We separate the running of a localization experiment from the visualization of its results.
We may run an experiment once to produce its log files and visualize those log files many times without repeating the experiment.

The mesh file and its required number of FLSs define the scale of an experiment, i.e., its number of cores.
We start by introducing small scale experiments involving point clouds that require 16 FLSs.
These introduce the experimentalist to the different point clouds considered in~\cite{swarical2024}, the use of dead reckoning that distorts a shape, and the visualization of the output files generated by experiments. We provide the log files from previously conducted thousand core experiments for visualization.
Section~\ref{sec:large} describes how an experimentalist may run such an experiment to produce the log files. 

\subsection{Small Scale Experiments}\label{sec:small}
Clone the Swarical repository:\\
\texttt{\small 
git clone https://github.com/flyinglightspeck/Swarical.git}\\
Open a browser to the Swarical GitHub link \url{https://github.com/flyinglightspeck/Swarical} and follow its README file.  
We recommend using Docker~\cite{docker2014} to reproduce the results of this section.
Depending on your operating system, 
install Docker Desktop or Engine.
Once installed, ensure Docker is running before proceeding.
From your desktop, open a terminal, change directory to the Swarical repository (\texttt{\small cd Swarical}), and execute its bash file (\texttt{\small bash init.sh}).
The latter builds the Docker image required for reproducibility.
It is simple to restart with a fresh environment by stopping and restarting the Docker daemon and executing the bash file.
Once the bash file execution completes, 
it will show:\\
\noindent \texttt{\small
To access the server, open this file in a browser:\\...\\http://127.0.0.1:8888/tree?token=e8b9aa5a6e22952673c053fdf}

\noindent The line starting with the \texttt{\small http://127.0.0.1:8888/...} will be different for your Docker instance.  Copy and paste it into a browser.
This opens an interface to the Jupyter Notebook running inside the Docker container.
Open the notebook named \texttt{\small reproduction.ipynb}.  
Click the play button at the top of Jupyter to step through the software.
The symbols next to Python software show whether the command is running (a *) or has completed execution (a number).  

The software starts by introducing seven different shapes (i.e., datasets) used in the experiments of~\cite{swarical2024}.
It instantiates the planner by specifying the radius of an FLS, and its camera’s minimum and maximum range.
The camera is the sensor an FLS uses to position itself relative to another FLS, see Section~\ref{camera}.
Next, we load the planner with one of the shapes by specifying its mesh file.
The planner reports the surface area of the mesh file in $meters^2$ and the number of required FLSs. Subsequently, it samples points from the mesh file and generates a point cloud. The user may visualize the resulting point cloud and how it may be illuminated with FLSs using dead reckoning~\cite{swarical2024}.
With this technique, an FLS travels from a dispatcher to its destination with a random pre-specified angle of error not to exceed \texttt{\small alpha}.
One may visualize the original point cloud, the ground truth, and the point cloud after the FLSs are dispatched to their assigned coordinate using dead reckoning.

Next, invoke the planner's compute\_trees method to compute a swarm-tree consisting of multiple FLS-trees.
The input is the approximate number of FLSs that should constitute a swarm.
One may save the resulting swarm tree in a file, insert dark FLSs to provide connectivity between its disjoint swarms (or isolated FLSs), and visualize the swarm tree.
All the above steps are fast and in total should not require more than one Second~\footnote{On a MacBook Pro with M1 processor.} to execute.
One may generate a swarm tree for all seven shapes and a variety of swarm sizes in the order of a second or two.

The next section of the notebook, Decentralized Localization, highlights Swarical's online localization with a small scale experiment that one may execute on a laptop.
It involves 16 FLSs (processes) illuminating a 2D rectangular array in a 4x4 grid.
First, it shows the planner computing the swarm-tree.
Next, it executes the localization technique.
By default, it will execute for 30 seconds.
It is computationally intensive for a small scale laptop or desktop.
The program consists of one primary process, resembling the Orchestrator~\cite{shahram2022}, and $F$=16 FLS processes.
The FLS processes execute the decentralized localization technique ISR~\cite{swarical2024}.
It requires FLSs to exchange messages and adjust their location continuously. 
We force it to terminate after 60 Seconds and to generate log files stored in the results directory. The files include the trajectory of each FLS and the messages they send and receive. The primary process uses these log files to compute the Hausdorff Distance and the Chamfer Distance as a function of time. These metrics show how closely the formation of FLSs matches the point cloud. Lower is better. The log files can be used to create a video visualization of FLSs moving to correct their position, converting the erroneous point cloud to the ground truth point cloud. FLSs that are members of the same swarm are shown with the same color.
Make sure to copy the log file produced by Line 16 and paste it for use in Line 17.
The process of creating the animation may require several minutes.
To view the animation, use the browser to navigate to the Docker directory containing it and download the mp4 file.
The Hausdorff distance is shown in the top left corner and drops below 1 cm in a few seconds.

The online experiments do not generate the exact same results each time due to the nondeterministic nature of their multitasking design. CPU scheduling, available resources, and the message passing delays vary the order in which the FLSs move, producing variation in the raw results, e.g, FLS trajectories. However, in each run, the results show Swarical is fast and quickly reduces Hausdorff and Chamfer Distances.

One may produce animations from an experiment's log files and view them unlimited times.
Step 21 illustrates this for log files from thousand core experiments conducted using CloudLab and Amazon AWS.
These experiments pertain to different variants of Swarical (HC, ISR, RSF) localizing the Skateboard. 
Their video generation may require tens of minutes.
They correspond to our original experiments reported in Figures 13, 14, and 15, and the videos presented in the caption of Figure 13 (\m{\href{https://youtu.be/GncnoqqYT_w}{ISR}}, \m{\href{https://youtu.be/0_Gs7IkDADw}{HC}}, and \m{\href{https://youtu.be/YlLCxW32tvg}{RSF}}). These files are automatically downloaded and placed in the results directory during the build process of the Docker image. A cell in the \texttt{reproduction.ipynb} is dedicated to generating the videos, and a Mathematica notebook, \texttt{swaricalplots.nb}, in the Swarical repository is provided to convert the raw results to the plots.
%We have provided the required instructions and scripts to run large-scale experiments on two platforms, Amazon AWS and CloudLab. 

\vspace*{-0.09in}\subsection{Decentralized Localization: Online Experiments}\label{sec:large}

This section describes how to conduct the experiments that produced the log files visualized in Section~\ref{sec:small}.
These were conducted using either 
CloudBank~\cite{cloudbank2021} or the Amazon AWS via CloudLab~\cite{emulab}.
Our Swarical repository provides detailed instructions for each, see
\ha{\url{https://github.com/flyinglightspeck/Swarical/blob/main/CloudLab_README.md}}
and
\ha{\url{https://github.com/flyinglightspeck/Swarical/blob/main/AWS_README.md}},
respectively. 
The choice of a cloud provider is not important because the results in~\cite{swarical2024} emphasize trends and comparisons instead of absolutes.
What is important is for a cluster dedicated to an experiment to have either the same number of cores or a few more cores than the number of FLSs $F$ required by illuminated shapes.
(Recall that the planner computes $F$ and reports its value for a shape, see Section~\ref{sec:small}.)

Our implementation uses the UDP broadcast protocol. It resembles FLSs in an area where they are communicating with one another.
This must be manually set up with Amazon AWS, and the README file provides instructions for it.
With the CloudLab, it is important for the PLATFORM in constants.py to be set to cloudlab.
Our scripts use this variable to configure a deployment to limit the broadcast to a group of servers in the cluster. Otherwise, a deployment will broadcast messages throughout a data center, causing inefficiency for CloudLab and its users\footnote{Such broadcasts are considered a violation.  Administrators may freeze the account of the violator and shut down the experiment.}.
The README file provides instructions. 

\subsection{Datasets}
We used five shapes from the Princeton Benchmark~\cite{princetonbenchmark} to evaluate Swarical: a chess piece, a dragon, a palm, a skateboard, and a racecar.
% They were converted into a collection of 3D coordinates for spherical FLSs using the technique described in~\cite{shahram2022}. 
Planner converts them into a collection of 3D coordinates for spherical FLSs using Poisson disk sampling\cite{6143943}.
With Chess, we include both the original and a down-sampled version. We also include a kangaroo mesh file from free3d.com website. These files are placed in \texttt{assets/dataset/mesh} directory.
Section~\ref{sec:small} described how these are registered with the planner.

\subsection{Reproducing Experimental Results}
Figures 1, 5, 10, 11, and 12 of~\cite{swarical2024} were generated using the planner.  
The final plots for Figures 10, 11, and 12 were created using the Wolfram Mathematica software. A Mathematica notebook called \texttt{swaricalplots.nb} converts raw results to the plots.

\section{Position Estimation Using a Camera}\label{camera}

In~\cite{swarical2024}, an FLS uses a Raspberry Pi Camera Module 3 NoIR\footnote{To function in the dark.} with autofocus and ArUco markers to compute its position relative to another FLS.
The relative state between two FLSs $u$ and $v$ includes a position $\relpose{u}{v}$ and an orientation (roll, pitch, and yaw).
Ideally, the accuracy of the position should be on the order of millimeters. 
The error in orientation should be less than 1° in each dimension.

\begin{figure}[ht]
\centering
\includegraphics[width=\columnwidth]
{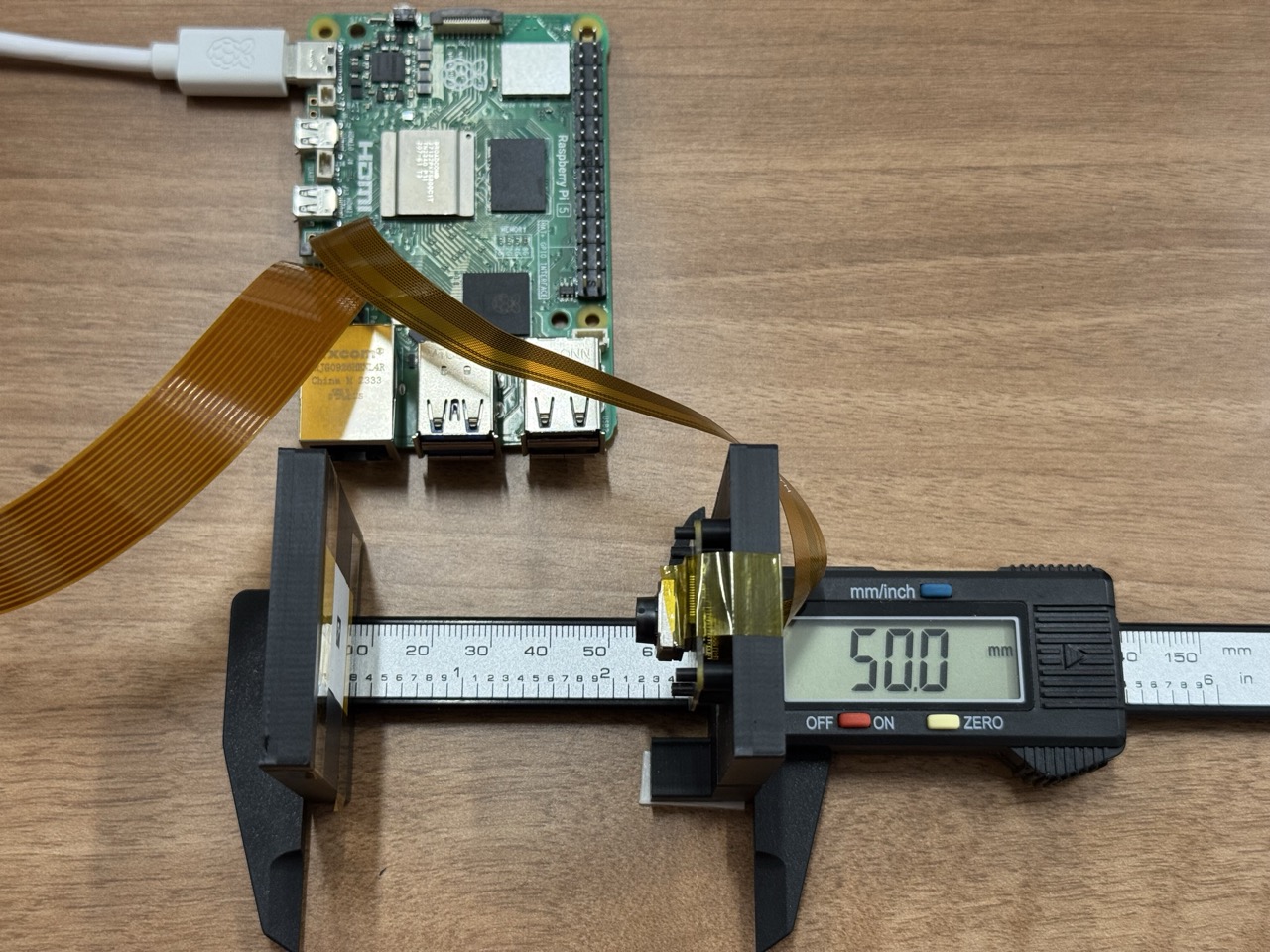}\hfill
\caption{Position estimation using printed paper.}
\label{fig:printed}
\end{figure}

The camera supports two types of lenses, regular and wide.  Figure 6 of~\cite{swarical2024} shows the blind, sweet, and decaying range of the camera~\cite{swazure2024} with each lens.
The camera does not detect an ArUco marker of a neighboring FLS when it is in its blind range, detects it with high accuracy when it is in its sweet range, and its detection becomes a function of distance when the marker is in the decaying range of the camera lens. 

We evaluated two possibilities for tagging an FLS with an ArUco marker:  1) printing the ArUco marker on paper and pasting it on each FLS, and 2) displaying it on an LCD mounted on an FLS.
Figure~\ref{fig:printed} shows our experimental setup with paper. It consists of:
\begin{itemize}
    \item Raspberry Pi 5 with Raspberry Pi OS (bookworm) installed. This component hosts the software that uses the output of the camera to quantify a relative position.
    \item 3D printed parts for the Raspberry Pi Camera Module 3 (Wide or Regular) and the holder for the printed paper. The files for the 3D models are available in \texttt{assets/paper}. 
    \item Caliper or ruler for measurements.
\end{itemize}

\begin{figure}[ht]
\centering
\includegraphics[width=\columnwidth]
{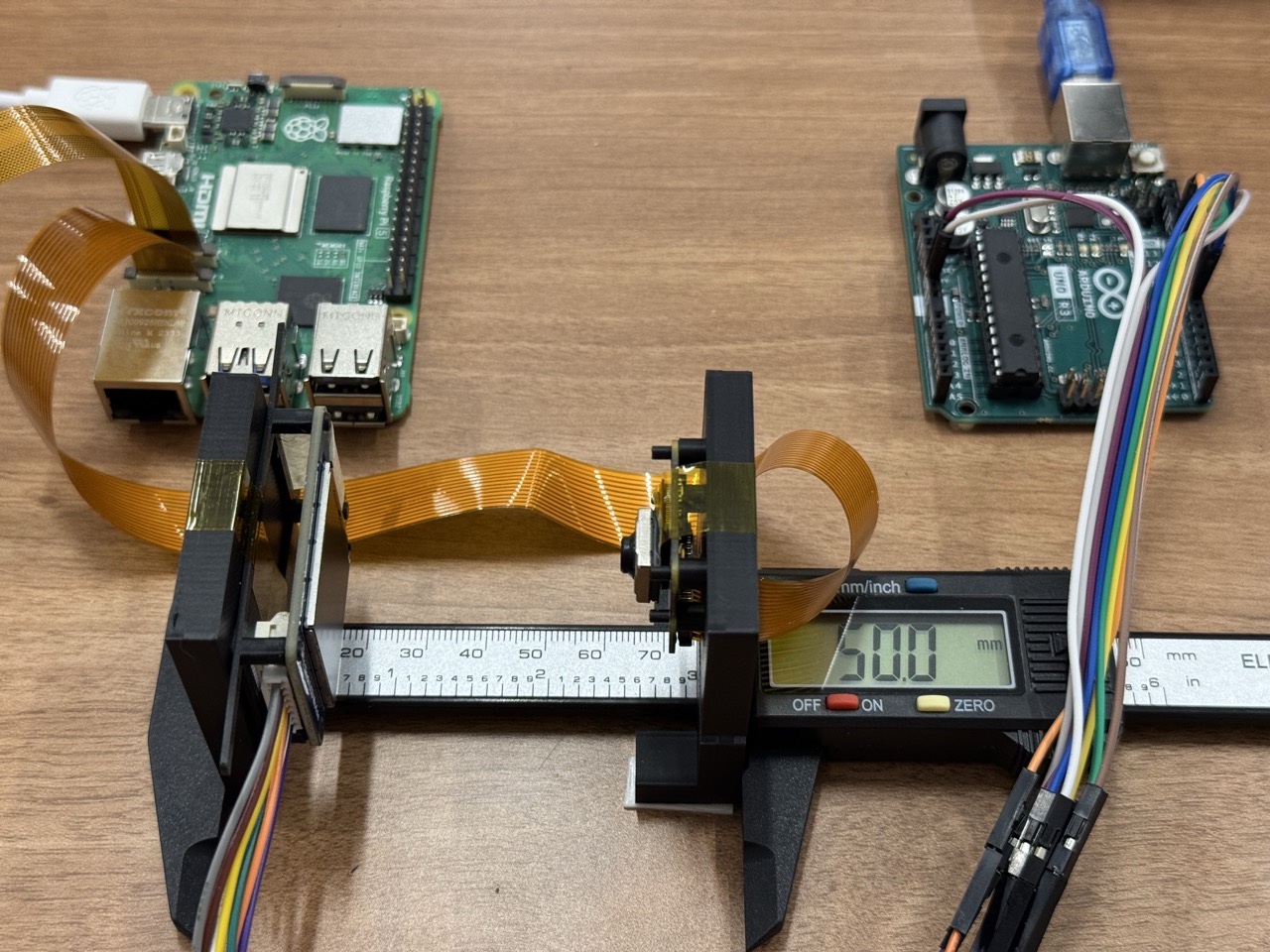}\hfill
\caption{Position estimation using an LCD display.}
\label{fig:lcd}
\end{figure}

Figure~\ref{fig:lcd} shows our setup with an LCD.
It consists of the same caliper, Raspberry Pi 5, and camera as Figure~\ref{fig:printed} with the addition of:
\begin{itemize}
    \item Waveshare 1.3 inch LCD Display Module IPS Screen 240x240.
    \item Arduino Uno Rev 3 to display the ArUco marker on the LCD.
    \item 3D printed parts for the Raspberry Pi Camera Module 3 (Wide or Regular), see \texttt{assets/lcd} for 3D printable models.  
\end{itemize}
% This setup quantifies the relative position vertically, while the setup of Figure~\ref{fig:printed} quantifies the same metric horizontally.

Figures 6 and 7 of~\cite{swarical2024} show the percentage error in the measured distances by the camera with both a wide and a regular lens using an ArUco marker on a printer paper with a 4.7 mm marker size.
The same setup is used to quantify the orientation error reported in Figure 9 of~\cite{swarical2024}.
Figure 8 of~\cite{swarical2024} highlights the difference between printing the ArUco marker on the paper versus displaying it on a small LCD.
For each, it shows the percentage error in measured distance with either the regular or wide angle lens.
Below, we describe how to reproduce these results.
(Also, see \ha{\url{https://github.com/flyinglightspeck/aruco-pose-estimation}} for the README file.)

With both setups, connect the camera to the Raspberry Pi (Pi for short) and turn it on.
Clone and install the requirements on the Pi:\\
\texttt{\small 
git clone \url{https://github.com/flyinglightspeck/aruco-pose-estimation.git} \\
cd aruco-pose-estimation \\
bash setup.sh}\\
Verify that the Pi detects the camera using the following commands:\\
\texttt{\small 
\hspace*{0.5in}rpicam-hello -{}-list-cameras \\
\hspace*{0.5in}rpicam-hello}\\
Initiate the virtual environment:\\
\texttt{\small 
\hspace*{0.5in}source .env/bin/activate}\\
Run the following command to start the program. It will run for 10 seconds and detect ArUco marker in each frame in real time:\\
\texttt{\small 
python pi\_pose\_estimation.py -i pi3 -r 720p -t 10 -{}-marker\_size 0.0047 -{}-live}\\
The user may save the results using the following command:\\
\texttt{\small 
python pi\_pose\_estimation.py -i pi3 -r 720p -t 10 -{}-marker\_size 0.0047 -{}-save -e 80mm}\\
The options \texttt{\small -i} and \texttt{\small -r} specify the camera type and its frame setting, respectively.  (See Table 2 of~\cite{swarical2024} for details.)  They utilize our provided calibration using the \texttt{\small calibration.py} file.
%Calibrate the camera with the 720p frame setting, see the \texttt{calibration} directory.
%\ha{We have already calibrated a total of 9 configurations using 3 cameras and 3 resolutions. The calibration results are stored in \texttt{calibration} directory and are ready to use for the program. If needed, you can repeat the calibration process with your camera. We used 720p resolution with the wide and regular cameras.}
The 720p setting was used for all experimental results reported in~\cite{swarical2024}.
%{\bf Hamed, provide instructions on how to do this caliberation and include in the README of the github repo.} \ha{Instructions on calibration are in the README file. Calibration results are already in the repository for a list of cameras and resolutions in the Calibration section of the README.}

The README of the repository enumerates the possible arguments of pi\_pose\_estimation.py.  By specifying the distance being measured using the \texttt{\small -e} option, the software computes and reports the observed error. This value should be set to the distance between the marker and the camera's image sensor. The 3d printed part has a placeholder for the camera's image sensor to facilitate adjusting the setup.
One may conduct many experiments with different distances to populate a \texttt{\small /results} directory with result files and batch process them by issuing \texttt{\small python process.py}.
This facilitates rapid data gathering with a final processing step.
The code to generate the plots used in Figures 6-9 of the paper is in the \texttt{\small cameraplots.nb} Mathematica notebook.

\section{Reproducibility Efforts}

Our reproducibility study encompassed three experimental setups, using the public codebase identified in Section~\ref{sec:intro}. 

First, we used Linux servers to re-execute the small-scale experiments responsible for the illustrative figures of~\cite{swarical2024}.
A subset of 3D meshes required minor, author-supplied code patches,
and an early race condition in the parallel execution code produced transient error logs until corrected.
Ultimately, all figures were successfully regenerated.
Second, the large-scale experiments were also reproduced without numerical deviation, but only after a protracted installation and configuration process on an Amazon AWS cloud instance with large distributed computational power; this stage would include non-negligible economic costs. 

Lastly, we re-conducted a subset of the experiments
to quantify the error in visual pose estimation using ArUco fiducial markers.
This took time because it required 
obtaining various hardware components,
printing 3D models to reconstruct the experimental setup,
and then re-conducting the experiment.
We used a Raspberry Pi camera module 3 with a wide-angle lens and
\emph{with} IR filter, whereas the original paper tested both wide-angle and regular lenses
-- both \emph{without} IR filters.
This is because the camera of the original paper was unavailable during the reproduction.
After clarifying several aspects of the experimental setup which were incorporated in this paper and the provided source code,
we reproduced 4 of 5 experiments successfully,
%done with the wide-angle lens camera in the original paper,
obtaining similar results with only small variations (likely due to manufacturing inconsistencies).
Sharing the process via video, in addition to reading the paper, was crucial to accurately converging on the
correct methodology.
Both the original draft of this paper and its online documentations have been updated to incorporate constructive comments provided by the reviewers.

\section{Conclusions}\label{conc}
This paper describes how to reproduce the results of the Swarical localization technique~\cite{swarical2024} using a Raspberry Pi Camera Module 3.
In caliberating this camera, one quickly discovers that it has a blind range.
Using its wide lens, it cannot detect distances shorter than 50 mm.
In~\cite{swazure2024}, we introduce Swazure.
It requires FLSs to cooperate by communicating with one another.  This cooperation enables those FLSs that are closer than 50 mm to compute their relative position accurately.  
Swazure is a building block of Swarical when FLSs are required to operate in the blind range of their positioning sensor.

\begin{acks}
This research was supported in part by the NSF grants IIS-2232382 and CMMI-2425754.
We gratefully acknowledge CloudBank~\cite{cloudbank2021} and CloudLab~\cite{emulab} for the use of their resources to enable all experimental results presented in~\cite{swarical2024} and their reproduction in this~paper.
\end{acks}

\balance

%\section{Experiments}
%\input{experimetns}

% \begin{acks}
% \end{acks}

\bibliographystyle{ACM-Reference-Format}
\bibliography{refs}

@inproceedings{decentralized2023,
author = {Alimohammadzadeh, Hamed and Culbertson, Heather and Ghandeharizadeh, Shahram},
title = {An Evaluation of Decentralized Group Formation Techniques for Flying Light Specks},
year = {2024},
isbn = {9798400702051},
publisher = {Association for Computing Machinery},
address = {New York, NY, USA},
url = {https://doi.org/10.1145/3595916.3626460},
doi = {10.1145/3595916.3626460},
booktitle = {Proceedings of the 5th ACM International Conference on Multimedia in Asia},
articleno = {84},
numpages = {7},
location = {Tainan, Taiwan},
series = {MMAsia '23}
}

@inproceedings{uavmm2025,
author = {Alimohammadzadeh, Hamed and Ghandeharizadeh, Shahram},
title = {{Illuminating English Letters Using a Flying Light Speck}},
year = {2025},  
publisher = {Association for Computing Machinery},
address = {New York, NY, USA},
url = {https://doi.org/10.1145/3728482.3757388},
doi = {10.1145/3728482.3757388},
booktitle = {Proceedings of the 3rd International Workshop on UAVs in Multimedia: Capturing the World from a New Perspective}, 
numpages = {5}, 
location = {Dublin, Ireland}, 
series = {UAVM '25} 
}

@inproceedings{mcgeKeynote2025,
author = {Ghandeharizadeh, Shahram},
title = {{Flying Light Specks: Dronevision, Holodecks and Spatial
Computing}},
year = {2025},  
publisher = {Association for Computing Machinery},
address = {New York, NY, USA},
url = {https://doi.org/10.1145/3746278.3759395},
doi = {10.1145/3746278.3759395},
booktitle = {Proceedings of the 3rd International Workshop on Multimedia
Content Generation and Evaluation: New Methods and Practice (McGE '25)}, 
numpages = {2}, 
location = {Dublin, Ireland}, 
series = {McGE '25} 
}

@Inproceedings{integrate2025,
  author =      {Nima Yazdani and Ghandeharizadeh, Shahram},
  title =       {{Integration of 3D FLS Displays with 3D Authoring Tools}},
  publisher =   "ACM Press",
  address =     "New York, NY",
  year =        {2025},
  isbn =        {79-8-4007-2051-2/2025/10},
  booktitle =   {Proceedings of the Third ACM International Workshop on Interactive Extended Reality},
  numpages = {8}, 
  location = {Dublin, Ireland}, 
  url = {https://doi.org/10.1145/3746269.3760418},
  doi = {10.1145/3746269.3760418},
  series = {IXR '25}
}

@inproceedings{cmpfls2023,
author = {Ghandeharizadeh, Shahram and Oria, Vincent},
title = {{Virtual Reality, Augmented Reality, Mixed Reality, Holograms and Holodecks}},
year = {2023}, 
publisher = {Mitra LLC}, 
address = {Los Angeles, CA, USA}, 
url = {https://doi.org/10.61981/ZFSH2304}, 
doi = {10.61981/ZFSH2304},
booktitle = {The First International Conference on Holodecks}, 
numpages = {3}, 
pages = {38--40},
location = {Los Angeles, California}, 
series = {Holodecks '23} 
}

@inproceedings{flightpatterns2023,
author = {Zhu, Shuqin and Ghandeharizadeh, Shahram}, 
title = {{Flight Patterns for Swarms of Drones}},
year = {2023}, 
publisher = {Mitra LLC}, 
address = {Los Angeles, CA, USA}, 
url = {https://doi.org/10.61981/ZFSH2303}, 
doi = {10.61981/ZFSH2303},
booktitle = {The First International Conference on Holodecks}, 
numpages = {5}, 
pages = {29--33},
location = {Los Angeles, California}, 
series = {Holodecks '23} 
}

@inproceedings{flshaptic23,
author = {Chen, Yang and Alimohammadzadeh, Hamed and Ghandeharizadeh, Shahram and Culbertson, Heather},
title = {{Towards Enabling Complex Touch-based Human-Drone Interaction.}},
year = {2023},
booktitle = {IROS Workshop on Human Multi-Robot Interaction},
location = {Detroit, USA}
}

@INPROCEEDINGS{shahram2022b,
  author={Ghandeharizadeh, Shahram and Garcia, Luis},
  booktitle={CHI 2022 Workshop on Novel Challenges of Safety, Security and Privacy in Extended Reality}, 
  title={{Safety in the Emerging Holodeck Applications}}, 
  month={April},
  year={2022}}

@inproceedings{reliability2024,
author = {Alimohammadzadeh, Hamed and Zhu, Shuqin and Bai, Jiadong and Ghandeharizadeh, Shahram},
title = {{Reliability Groups with Standby Flying Light Specks}},
year = {2024},
isbn = {9798400704123},
publisher = {Association for Computing Machinery},
address = {New York, NY, USA},
url = {https://doi.org/10.1145/3625468.3647606},
doi = {10.1145/3625468.3647606},
booktitle = {Proceedings of the 15th ACM Multimedia Systems Conference},
pages = {1–11},
numpages = {11},
location = {Bari, Italy},
series = {MMSys '24}
}

@INPROCEEDINGS{imeta2023,
  author={Phan, Trung and Alimohammadzadeh, Hamed and Culbertson, Heather and Ghandeharizadeh, Shahram},
  booktitle={2023 International Conference on Intelligent Metaverse Technologies \& Applications (iMETA)}, 
  title={An Evaluation of Three Distance Measurement Technologies for Flying Light Specks*}, 
  year={2023},
  volume={},
  number={},
  pages={1-8},
  doi={10.1109/iMETA59369.2023.10294597}}

@inproceedings{mmsys2023,
author = {Alimohammadzadeh, Hamed and Mehraban, Daryon and Ghandeharizadeh, Shahram},
title = {{Modeling Illumination Data with Flying Light Specks}},
year = {2023},
pages={363–368},
publisher = {Association for Computing Machinery},
address = {New York, NY, USA},
doi = {https://doi.org/10.1145/3587819.3592544},
booktitle = {ACM Multimedia Systems},
location = {Vancouver, Canada},
series = {MMSys '23}
}

@inproceedings{integrate2023,
author = {Nima Yazdani and Alimohammadzadeh, Hamed and Ghandeharizadeh, Shahram},
title = {{A Conceptual Model of Intelligent Multimedia Data Rendered using Flying Light Specks}},
year = {2023}, 
publisher = {Mitra LLC}, 
address = {Los Angeles, CA, USA}, 
url = {https://doi.org/10.61981/ZFSH2309}, 
doi = {10.61981/ZFSH2309},
booktitle = {The First International Conference on Holodecks}, 
numpages = {7}, 
pages = {38--44},
location = {Los Angeles, California}, 
series = {Holodecks '23} 
}

@inproceedings{emulab,
 author = {White, Brian and Lepreau, Jay and Stoller, Leigh and Ricci, Robert and Guruprasad, Shashi and Newbold, Mac and Hibler, Mike and Barb, Chad and Joglekar, Abhijeet},
 title = {{An Integrated Experimental Environment for Distributed Systems and Networks}},
 journal = {SIGOPS Oper. Syst. Rev.},
 issue_date = {Winter 2002},
 volume = {36},
 number = {SI},
 month = dec,
 year = {2002},
 issn = {0163-5980},
 pages = {255--270},
 numpages = {16},
 url = {http://doi.acm.org/10.1145/844128.844152},
 doi = {10.1145/844128.844152},
 acmid = {844152},
 publisher = {ACM},
 address = {New York, NY, USA},
}

@inproceedings{cloudbank2021,
author = {Norman, Michael and Kellen, Vince and Smallen, Shava and DeMeulle, Brian and Strande, Shawn and Lazowska, Ed and Alterman, Naomi and Fatland, Rob and Stone, Sarah and Tan, Amanda and Yelick, Katherine and Van Dusen, Eric and Mitchell, James},
title = {{CloudBank: Managed Services to Simplify Cloud Access for Computer Science Research and Education}},
year = {2021},
isbn = {9781450382922},
publisher = {Association for Computing Machinery},
address = {New York, NY, USA},
url = {https://doi.org/10.1145/3437359.3465586},
doi = {10.1145/3437359.3465586},
booktitle = {Practice and Experience in Advanced Research Computing},
articleno = {45},
numpages = {4},
location = {Boston, MA, USA},
series = {PEARC '21}
}

@inproceedings{princetonbenchmark,
  author    = {Philip Shilane and
               Patrick Min and
               Michael M. Kazhdan and
               Thomas A. Funkhouser},
  title     = {{The Princeton Shape Benchmark}},
  booktitle = {2004 International Conference on Shape Modeling and Applications {(SMI}
               2004), 7-9 June 2004, Genova, Italy},
  pages     = {167--178},
  publisher = {{IEEE} Computer Society},
  year      = {2004},
  url       = {https://doi.org/10.1109/SMI.2004.1314504},
  doi       = {10.1109/SMI.2004.1314504},
  bibsource = {dblp computer science bibliography, https://dblp.org}
}

@article{standbyfls2025,
author = {Alimohammadzadeh, Hamed and Zhu, Shuqin and Ghandeharizadeh, Shahram},
title = {{Techniques to Conceal Dark Standby Flying Light Specks}},
year = {2025},
publisher = {Association for Computing Machinery},
address = {New York, NY, USA},
issn = {1551-6857},
url = {https://doi.org/10.1145/3724399},
doi = {10.1145/3724399},
note = {Just Accepted},
journal = {ACM Trans. Multimedia Comput. Commun. Appl.},
month = apr
}

@inproceedings{swarmer2023,
author = {Alimohammadzadeh, Hamed and Ghandeharizadeh, Shahram}, 
title = {{SwarMer: A Decentralized Localization Framework for Flying Light Specks}},
year = {2023}, 
publisher = {Mitra LLC}, 
address = {Los Angeles, CA, USA}, 
url = {https://doi.org/10.61981/ZFSH2302}, 
doi = {10.61981/ZFSH2302},
booktitle = {The First International Conference on Holodecks}, 
numpages = {13}, 
pages = {10--22},
location = {Los Angeles, California}, 
series = {Holodecks '23} 
}

@inproceedings{swarical2024,
author = {Alimohammadzadeh, Hamed and Ghandeharizadeh, Shahram},
title = {Swarical: An Integrated Hierarchical Approach to Localizing Flying Light Specks},
year = {2024},
isbn = {9798400706868},
publisher = {Association for Computing Machinery},
address = {New York, NY, USA},
url = {https://doi.org/10.1145/3664647.3681080},
doi = {10.1145/3664647.3681080},
booktitle = {Proceedings of the 32nd ACM International Conference on Multimedia},
pages = {6153–6161},
numpages = {9},
location = {Melbourne VIC, Australia},
series = {MM '24}
}

@inproceedings{flshaptics2023,
author = {Chen, Yang and Alimohammadzadeh, Hamed and Culbertson, Heather and Ghandeharizadeh, Shahram},
title = {{Towards a Stable 3D Physical Human-Drone Interaction}},
year = {2023}, 
publisher = {Mitra LLC}, 
address = {Los Angeles, CA, USA}, 
url = {https://doi.org/10.61981/ZFSH2308}, 
doi = {10.61981/ZFSH2308},
booktitle = {The First International Conference on Holodecks}, 
numpages = {4}, 
pages = {34--37},
location = {Los Angeles, California}, 
series = {Holodecks '23} 
}

@INPROCEEDINGS{haptics2024,
  author={Chen, Yang and Alimohammadzadeh, Hamed and Ghandeharizadeh, Shahram and Culbertson, Heather},
  booktitle={2024 IEEE Haptics Symposium (HAPTICS)}, 
  title={Force-Feedback Through Touch-based Interactions With A Nanocopter}, 
  year={2024},
  volume={},
  number={},
  pages={271-277},
  doi={10.1109/HAPTICS59260.2024.10520851}}

@inproceedings{circular2024,
author = {Zhu, Shuqin and Ghandeharizadeh, Shahram},
title = {{Circular Flight Patterns for Dronevision}},
year = {2024}, 
publisher = {Mitra LLC}, 
address = {Los Angeles, CA, USA}, 
url = {https://doi.org/10.61981/ZFSH2404}, 
doi = {10.61981/ZFSH2404},
booktitle = {The Second International Conference on Holodecks}, 
numpages = {11}, 
pages = {1--11},
location = {Los Angeles, California}, 
series = {Holodecks '24} 
}

@inproceedings{swazure2024,
author = {Alimohammadzadeh, Hamed and Ghandeharizadeh, Shahram},
title = {{Swazure: Swarm Measurement of Pose for Flying Light Specks}},
year = {2024}, 
publisher = {Mitra LLC}, 
address = {Los Angeles, CA, USA}, 
url = {https://doi.org/10.61981/ZFSH2403}, 
doi = {10.61981/ZFSH2403},
booktitle = {The Second International Conference on Holodecks}, 
numpages = {8}, 
pages = {17--25},
location = {Los Angeles, California}, 
series = {Holodecks '24} 
}

@Inproceedings{shahram2021,
  author=       {Ghandeharizadeh, Shahram},
  booktitle=    {ACM Multimedia Asia}, 
  title=        {{Holodeck: Immersive 3D Displays Using Swarms of Flying Light Specks}}, 
  month=        {December},
  publisher =   "ACM Press",
  address =     "New York, NY",
  pages =       {1-7},
  year=         {2021},
  location =    {Gold Coast, Australia},
  doi=          {https://doi.org/10.1145/3469877.3493698}
}

@inproceedings{shahram2022,
author = {Ghandeharizadeh, Shahram},
title = {Display of 3D Illuminations using Flying Light Specks},
year = {2022},
isbn = {9781450392037},
publisher = {Association for Computing Machinery},
address = {New York, NY, USA},
url = {https://doi.org/10.1145/3503161.3548250},
doi = {10.1145/3503161.3548250},
booktitle = {Proceedings of the 30th ACM International Conference on Multimedia},
pages = {2996–3005},
numpages = {10},
location = {Lisboa, Portugal},
series = {MM '22}
}

@inproceedings{dv2023,
author = {Alimohammadzadeh, Hamed and Bernard, Rohit and Chen, Yang and Phan, Trung and Singh, Prashant and Zhu, Shuqin and Culbertson, Heather and Ghandeharizadeh, Shahram},
title = {{Dronevision: An Experimental 3D Testbed for Flying Light Specks}},
year = {2023}, 
publisher = {Mitra LLC}, 
address = {Los Angeles, CA, USA}, 
url = {https://doi.org/10.61981/ZFSH2301}, 
doi = {10.61981/ZFSH2301},
booktitle = {The First International Conference on Holodecks}, 
numpages = {9}, 
pages = {1--9},
location = {Los Angeles, California}, 
series = {Holodecks '23} 
}

@ARTICLE{6143943,
  author={Corsini, Massimiliano and Cignoni, Paolo and Scopigno, Roberto},
  journal={IEEE Transactions on Visualization and Computer Graphics}, 
  title={Efficient and Flexible Sampling with Blue Noise Properties of Triangular Meshes}, 
  year={2012},
  volume={18},
  number={6},
  pages={914-924},
  doi={10.1109/TVCG.2012.34}}

@article{docker2014,
author = {Merkel, Dirk},
title = {Docker: Lightweight Linux Containers for Consistent Development and Deployment},
year = {2014},
issue_date = {March 2014},
publisher = {Belltown Media},
address = {Houston, TX},
volume = {2014},
number = {239},
issn = {1075-3583},
journal = {Linux J.},
month = mar,
articleno = {2}
}

% \appendix

\end{document}